\documentclass[12pt]{article}
\usepackage{amsfonts}
\usepackage{mathrsfs}
\usepackage{amsmath,amssymb,array}
\usepackage{graphicx}

\usepackage[top=3cm, bottom=3.2cm, left=2.5cm, right=2.5cm]{geometry}
\numberwithin{equation}{section}
\def\p{\partial}

\begin{document}

\begin{titlepage}
\renewcommand{\thefootnote}{\fnsymbol{footnote}}

\begin{center}

\begin{flushright}\end{flushright}
\vspace{2.0cm}

\textbf{\Large{Path Integral and Asian Options}}\vspace{2cm}

\textbf{Peng Zhang} \\[0.5cm]


\emph{Institute of Theoretical Physics, College of Applied Sciences, \\
      Beijing University of Technology, Beijing 100124, P.R.China}

\end{center}\vspace{1.5cm}

\centerline{\textbf{Abstract}}\vspace{0.5cm} In this paper we
analytically study the problem of pricing an arithmetically averaged Asian
option in the path integral formalism. By a trick about the Dirac
delta function, the measure of the path integral is defined by an
effective action functional whose potential term is an exponential function.
This path integral is evaluated by use of the Feynman-Kac theorem.
After working out some auxiliary integrations involving Bessel and
Whittaker functions, we arrive at the spectral expansion for
the value of Asian options.

\vspace{1.5cm}
August 2010.

\vspace{0.2cm}
Last revised: October 2013. 


\end{titlepage}
\setcounter{footnote}{0}

\section{Introduction}

It has been known for a long time that the path integral formalism can be applied to
the pricing of financial securities. The standard methods in quantitative finance are
the stochastic calculus and partial differential equations. In almost all cases the relevant differential
equations are diffusion type, whose solution is determined by the heat kernel. It is well-known
that \cite{FH65} the heat kernel can be written in terms of a path integral. Sometimes this fact is called Feynman-Kac theorem.
This is the starting point for the financial applications of the path integral formalism.
In \cite{D88, D89} it has been applied to the European options and the one-factor term-structure models.
In \cite {L98, M00} it is shown how the pricing of path-dependent options can be incorporated into the
path integral formulation. It is applied in \cite{B97} to models with stochastic volatility, and in \cite{B98}
by the same author to the Heath-Jarrow-Morton model of forward interest rates. See e.g. \cite{O98}-\cite{B04}
for more works in this direction.

Among many exotic options in the financial market, the asian option is a very popular one.
Its payoff depends on the arithmetic average of the price of the underling asset during the
life of this option contract. The Asian options has the advantage that it is usually less expensive
than standard options due to its smaller volatility, and its value is harder to be manipulated by
a large market participant. So it is more safe to hold it. On the theoretical side, the exact
pricing of the arithmetically averaged Asian option is a challenging problem, since the arithmetical
average of a stochastic variable, which is logarithmic-normal, is not logarithmic-normally distributed
anymore. In the pioneering work \cite{GY93}, Geman and Yor derived a closed form expression for the
Laplace transformation of the value of the Asian option. In \cite{L04} Linetsky obtained a spectral
expansion expression of its value in terms of confluent hypergeometric functions. This work is the main
motivation of the present paper.

In this paper we will study the the arithmetically averaged Asian option in the path integral
formalism. In section 2, we review the path integral formulation for a general
path-dependent option. For the Asian option, the resulting effective action is the (imaginary-time)
quantum mechanics with an exponential potential energy. In section 3, we obtain the corresponding heat kernel by solving
the differential equation it satisfies. We work out two auxiliary integrals in the following section 4 and 5.
In section 6, we specify the payoff function and study the value of the put and call options. For the
put option we can directly calculate its value, which is equivalent to the result of \cite{L04},
while for the call option we can use the put-call parity relation.

\section{Path integral formulation}

If $S$ denotes the price of a stock, it is commonly assumed that $X:=\log{S}$ is
a Brownian motion. A stock option is a kind of financial derivative whose value $\mathcal{O}$ depends
on the behavior of the underlining stock price \cite{H08}. For a general path-dependent option, its
payoff $\Phi[X]$ is a functional of $X$. That means the final payoff of this option contract
depends on the whole history of the stock price before its maturity. The usual European option is
just a special case, whose payoff functional is local, i.e. only depends on the stock price at the
maturity day.

By the risk-neutral pricing formula,
the value of a 
stock option can be written in a path integral form as (see e.g. \cite{L98})
\footnote{When $\Phi[X]=(e^{X(T)}-K)^+$, it leads to the classic Black-Scholes formula.}
\begin{equation*}
\mathcal{O}=\,e^{-r(T-t)}\,\tilde{\mathbb{E}}[\Phi]=\,e^{-r(T-t)}\int_{-\infty}^{\infty}dx' \int_{X(t)=x\,\, \atop X(T)=x'}\hspace{-0.15cm}DX
                \exp \left\{\frac{1}{2\sigma^2}\int_{\,t}^{T}\left(-\frac{dX}{dt'}+\mu\right)^2dt'\right\}\, \Phi[X].
\end{equation*}
In the above equation, $\int DX$ means the formal integration over all paths $X$ with $X(t)=x$ and $X(T)=x'$.
In addition, $r$ denotes the risk-free interest rate, $\sigma$ is the volatility of the stock
price, and $\mu:=r-\sigma^2/2$. All of these parameters are assumed to be constant.
By introducing the following combinations
\begin{eqnarray}
R=\frac{r}{\sigma^2}\,,\quad
\tau=\sigma^2 (T-t)\,,\quad
\nu=\frac{2\mu}{\sigma^2}\,,
\end{eqnarray}
we can simplify the above expression of $\mathcal{O}$ as
\begin{eqnarray}
\mathcal{O}(\tau,x)=\,\,e^{-R\tau}\int_{-\infty}^{\infty}dx'\,e^{\nu(x'-x)/2-\nu^2\tau/8}\int_{X(\tau)=x\, \atop X(0)=x'}DX\,
                \exp \left\{-\frac{1}{2}\int_0^\tau \dot{X}^2\,d\tau'\right\}\, \Phi[X]\,.\label{O1}
\end{eqnarray}
Suppose the payoff functional takes the form $\Phi[X]=\phi\,(V[X])$ for some function $\phi$, with
\begin{eqnarray}
V=\,V[X]=\,\,\tau^{-1}\int_0^\tau e^X d\tau'\,. \label{VV}
\end{eqnarray}
For example, the put Asian option has $\phi_P(V)=(K-V)^+$, while the call one has $\phi_C(V)=(V-K)^+$.
Then we have 
\begin{align}
\Phi[X]\, &=\,\phi(V)\,=\,\,\int_{0}^\infty\delta(\xi-V)\,\phi(\xi)\,d\xi \nonumber\\
          &=\,\int_0^\infty d\xi\,\,\phi(\xi)\times\,\frac{1}{2\pi i}
                     \int_{\epsilon-i\infty}^{\epsilon+i\infty}\,e^{\,q(\xi-V)}\,dq\,.\label{F}
\end{align}
We can restrict the integration range to $(0,\infty)$ in the first line because the functional
$V=V[X]$ is always positive. In the second line we have use the Laplace transformation of the
Dirac delta function with $\epsilon$ being any positive real number. Insert (\ref{VV}) and (\ref{F}) into
(\ref{O1}), followed by rescaling some integration variables, we have
\begin{eqnarray}
\mathcal{O}(\tau,x)&=&\frac{e^{-R\tau}}{2\pi i}\,\,e^{-\nu x/2-\nu^2\tau/8}\,
            \int_0^\infty d\xi\,\,\phi(\xi/\tau)\,\int_{\epsilon-i\infty}^{\epsilon+i\infty}dq\,\,e^{q\xi}  \nonumber\\[0.2cm]
&&\hspace{0.5cm}
            \times\,\int_{-\infty}^{\infty}dx'\,e^{\nu x'/2}
            \int_{X(\tau)=x\, \atop X(0)=x'}DX\,\exp\left\{-\int_0^\tau\left(\frac{1}{2}\,\dot{X}^2+q\,e^X\right)d\tau'\right\}\,.\label{O2}
\end{eqnarray}
This is our path integral formulation of the valuation of the Asian option.
We see that the system is driven by an effective action
\begin{eqnarray}
A_{\mathrm{eff}}[X]\,=\,\int_0^\tau\left(\frac{1}{2}\,\dot{X}^2+q\,e^X\right)d\tau'\,,\label{Aeff}
\end{eqnarray}
which is called (imaginary-time) Liouville quantum mechanics in \cite{DJ82}.

\section{Heat kernel}

Define the heat kernel
\begin{eqnarray}
\mathcal{K}(\tau,x,x';q)\,:=\,\int_{X(\tau)=x\, \atop X(0)=x'}DX\,\exp\left\{-\int_0^\tau\left(\frac{1}{2}\dot{X}^2+q\,e^X\right)d\tau'\right\}\,.
\end{eqnarray}
By use of Feynman-Kac theorem, it satisfies the following initial value problem
\begin{eqnarray}
-\frac{\p \mathcal{K}}{\p \tau}&=&-\frac{1}{2}\,\frac{\p^2\mathcal{K}}{\p x^2}+q\,e^x \mathcal{K}\,,\nonumber \\[0.3cm]
\mathcal{K}|_{\tau=0}&=&\delta(x-x')\,. \label{EqK}
\end{eqnarray}
Actually this is the (imaginary-time) Schr\"{o}dinger equation of the effective action (\ref{Aeff}).
We may use the method of spectral expansion to construct the heat kernel $\mathcal{K}(\tau,x,x';q)$.
Firstly we solve the following eigenvalue/eigenfunction problem
\begin{eqnarray}
-\frac{1}{2}\,\frac{\p^2\psi_u}{\p x^2}+q\,e^x \psi_u=\frac{u^2}{8}\,\,\psi_u\,.
\end{eqnarray}
For $q\in\mathbb{C}$ with $|\mathrm{arg}\,q|<\pi$, the normalized eigenfunction is
\begin{eqnarray}
\psi_u(x)\,=\,\frac{1}{\pi}\sqrt{u\sinh(\pi u)}\,\,K_{iu}(\sqrt{8q}\,e^{x/2})\,,\qquad u>0\,, \label{EigFn}
\end{eqnarray}
where $K_{iu}$ is the modified Bessel function of the second kind.
Then the heat kernel can be written as
\begin{eqnarray}
\mathcal{K}(\tau,x,x';q)&=&\int_0^\infty e^{-u^2\tau/8}\,\psi_u(x)\,\psi_u(x')\,du \nonumber \\[0.2cm]
        &=&\frac{1}{\pi^2}\int_0^{\infty}e^{-u^2\tau/8}K_{iu}(\sqrt{8q}\,e^{x/2})\,K_{iu}(\sqrt{8q}\,e^{x'/2})\sinh(\pi u)\,u\,du\,. \label{K}
\end{eqnarray}
It can be explicitly proved that, when $q>0$, (\ref{K}) is indeed the solution of (\ref{EqK}).
That it satisfies the differential equation can be easily checked. In the appendix we will
show that it also satisfies the initial condition, i.e. the completeness of $\{\psi_{iu}(x)\,|u>0\}$.
Due to the inverse Laplace transformation in (\ref{O2}), we need $q$ to be complex with $\mathrm{Re}\,q>0$.
Since the solution of (\ref{EqK}) should be a holomorphic function in the complex $q$-plane cut open along
the negative real axis, we can insert (\ref{K}) into (\ref{O2}) to calculate the option value, and the result
turns out to be correct. Therefore the expression of the option value becomes
\begin{eqnarray}
\mathcal{O}(\tau,x)&=&\frac{e^{-R\tau}}{2\pi i}\,\,e^{-\nu x/2-\nu^2\tau/8}
   \int_0^\infty d\xi\,\phi(\xi/\tau)\int_{\epsilon-i\infty}^{\epsilon+i\infty}dq\,\,e^{q\xi}
   \int_{-\infty}^{\infty}dx'\,e^{\nu x'/2}  \nonumber\\[0.2cm]
& &\hspace{0.5cm}
   \times\,\frac{1}{\pi^2}\int_0^{\infty}e^{-u^2\tau/8}K_{iu}(\sqrt{8q}\,e^{x/2})\,K_{iu}(\sqrt{8q}\,e^{x'/2})\sinh(\pi u)\,u\,du\,.\label{O3}
\end{eqnarray}

\section{Integrating out $x'$}

In this section we will consider the integration over the variable $x'$. Define
\begin{eqnarray}
\mathcal{M}(\tau,x\,;q):=\int_{-\infty}^{\infty}e^{\nu x'/2}\,\mathcal{K}(\tau,x,x';q)\,dx'\,,
\end{eqnarray}
which satisfies the following initial value problem
\begin{eqnarray}
-\frac{\p\mathcal{M}}{\p \tau}&=&-\frac{1}{2}\,\frac{\p^2\mathcal{M}}{\p x^2}+q\,e^x\mathcal{M}\,,\nonumber \\[0.3cm]
\mathcal{M}\,|_{\tau=0}&=&e^{\nu x/2}\,. \label{Eqf}
\end{eqnarray}
To solve this problem we use the following expansion of the initial configuration \cite{PP10}
\begin{eqnarray}
e^{\nu x/2}&=&\frac{2^{-1-\nu/2}}{\pi^2\,q^{\nu/2}}\int_0^{\infty}\,
     \left|\,\Gamma\left(\frac{\nu+iu}{2}\right)\,\right|^{\,2}K_{iu}(\sqrt{8q}\,e^{x/2})\,\sinh(\pi u)\,u\,\,du \nonumber\\[0.2cm]
&&\hspace{0.5cm}+\,\frac{2^{1-\nu/2}}{q^{\nu/2}}\sum_{n=0}^{[-\nu/2\,]}\frac{(-\nu-2n)}{n!\,\Gamma(-\nu-n+1)}\,K_{-\nu-2n}(\sqrt{8q}\,e^{x/2})\,.
\end{eqnarray}
Note that when $\nu$ is not positive, the function $e^{\nu x/2}$ is not in $L^2(\mathbb{R})$, so the above equation
cannot be argued by just the orthogonality. Since eigenfunctions evolve independently, the solution of (\ref{Eqf}) is
\begin{eqnarray}
\mathcal{M}(\tau,x\,;q)&=&\frac{2^{-1-\nu/2}}{\pi^2\,q^{\nu/2}}\int_0^{\infty}e^{-u^2\tau/8}\left|\,\Gamma\left(\frac{\nu+iu}{2}\right)\,\right|^{\,2}
            K_{iu}(\sqrt{8q}\,e^{x/2})\,\sinh(\pi u)\,u\,du \nonumber\\[0.2cm]
&& \hspace{0.2cm}+\,\frac{2^{1-\nu/2}}{q^{\nu/2}}\sum_{n=0}^{[-\nu/2\,]}\frac{(-\nu-2n)}{n!\,\Gamma(-\nu-n+1)}\,
            e^{(\nu+2n)^2\tau/8}\,K_{-\nu-2n}(\sqrt{8q}\,e^{x/2})\,.
\end{eqnarray}
Therefore the option value can be written as
\begin{eqnarray}
\mathcal{O}(\tau,x)&=&\frac{e^{-R\tau}}{2\pi i}\,\,e^{-\nu x/2-\nu^2\tau/8}\int_0^\infty d\xi\,\phi(\xi/\tau)
            \int_{\epsilon-i\infty}^{\epsilon+i\infty}dq\,\,e^{\,\xi q}\mathcal{M}(\tau,x\,;q)\,.
\end{eqnarray}
Let us define the pricing kernel $\mathcal{P}(\tau,x,\xi)$ by
\begin{eqnarray}
\mathcal{P}(\tau,x,\xi)&:=&\frac{1}{2\pi i}\,\int_{\epsilon-i\infty}^{\epsilon+i\infty}
         4\,e^{\,\xi q}\,e^{(1-\nu/2)x-\nu^2\tau/8}\mathcal{M}(\tau,x\,;q)\,dq\\[0.2cm]
&=&\frac{2^{1-\nu/2}}{\pi^2}\,\,e^{(1-\nu/2)x}\int_0^\infty du\,\sinh(\pi u)\,u \,
         e^{-(u^2+\nu^2)\tau/8}\left|\,\Gamma\left(\frac{\nu+iu}{2}\right)\,\right|^{\,2}  \nonumber\\[0.2cm]
&&\hspace{1cm}\times\,\,\frac{1}{2\pi i}\,\int_{\epsilon-i\infty}^{\epsilon+i\infty}\,e^{\,\xi q}\,q^{-\nu/2}K_{iu}(\sqrt{8q}\,e^{x/2})\,dq \nonumber\\[0.3cm]
&& +\,2^{3-\nu/2}\,e^{(1-\nu/2)x}\sum_{n=0}^{[-\nu/2\,]}\frac{(-\nu-2n)}{n!\,\Gamma(-\nu-n+1)}\,\,e^{n(\nu+n)\tau/2} \nonumber\\[0.2cm]
&&\hspace{1cm}\times\,\,\frac{1}{2\pi i}\,\int_{\epsilon-i\infty}^{\epsilon+i\infty}\,e^{\,\xi q}\,q^{-\nu/2}K_{-\nu-2n}(\sqrt{8q}\,e^{x/2})\,dq\,. \label{O4}
\end{eqnarray}
Then the option value becomes
\begin{eqnarray}
\mathcal{O}(\tau,x)&=&\frac{e^{-R\tau}}{\,4\, e^x}\,\int_0^{\infty}\mathcal{P}(\tau,x,\xi)\,\phi(\xi/\tau)\,d\xi\,.\label{O}
\end{eqnarray}
From this formula we can see that $\mathcal{P}(\tau,x,\xi)$ is essentially the probability density transition function
of the stochastic process $V_\tau:=\tau^{-1}\hspace{-0.1cm}\int_0^\tau e^X d\tau$.

\section{Integrating out $q$}

\begin{figure}
  \centering
  \includegraphics[width=12cm]{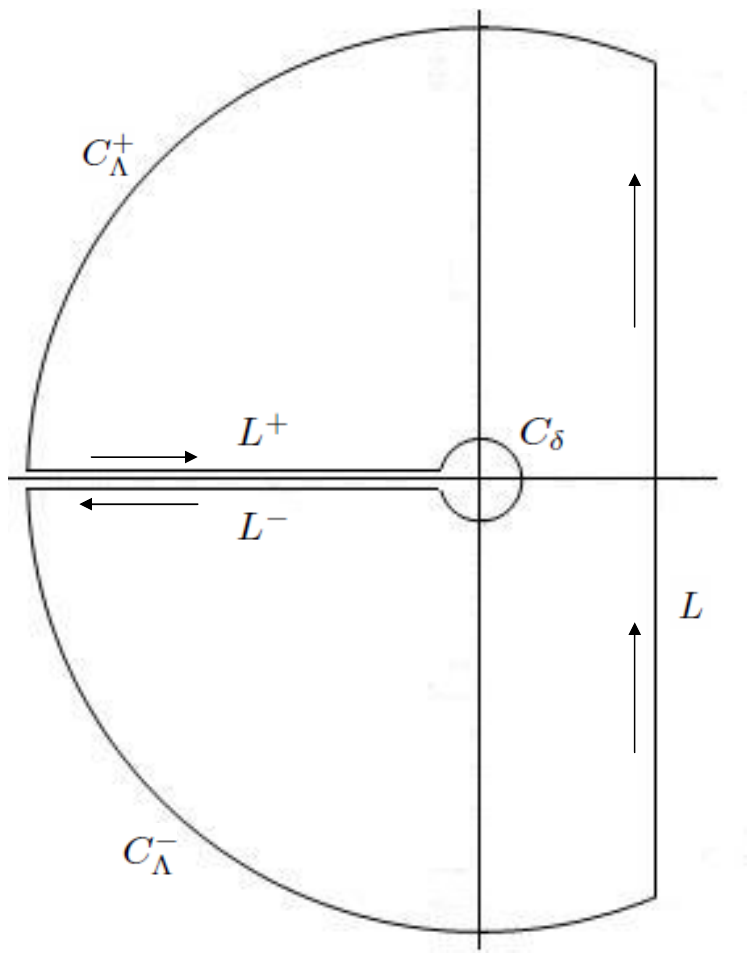} 
  \caption{\small{The integration contour for the calculation of the inverse Laplace transformation (\ref{InvL}).
         The radius of $C_\Lambda^{\pm}$ and $C_\delta$ are $\Lambda$ and $\delta$, respectively.}}\label{cont}
\end{figure}

In this section we will work out the inverse Laplace transformation in (\ref{O4}).
For this we consider the following integration
\begin{eqnarray}
I&:=&\frac{1}{2\pi i}\,\int_{\epsilon-i\infty}^{\epsilon+i\infty}\,e^{\,\xi q}\,q^{-\nu/2}K_\rho(\sqrt{8q}\,e^{x/2})\,dq\,. \label{InvL}
\end{eqnarray}
To calculate this inverse Laplace transformation, we use the contour as in Figure \ref{cont}.
It can be shown that the integration along $C_\Lambda^{\pm}$ tend to zero as $\Lambda\rightarrow\infty$.
When $\nu+|\mathrm{Re}\,\rho|<2$ the integration along $C_\delta$ also vanishes as $\delta\rightarrow0$.
Therefore the original integration along $L$ is related to the integration along $L^+$ and $L^-$.
Nevertheless it can be checked that the result we obtain in this way is still true for more general $\nu$.
Note that due to the multivaluedness of the integrand, its values along $L^+$ and $L^-$ are not same. We choose
$\mathrm{arg}\,q=\pi$ on $L^+$, while $\mathrm{arg}\,q=-\pi$ on $L^-$.
Explicitly we have
\begin{eqnarray}
I&=&-\frac{1}{2\pi i}\,\left(\int_{L^+}+\int_{L^-}\right)\,e^{\,\xi q}\,q^{-\nu/2}K_\rho(\sqrt{8q}\,e^{x/2})\,dq \nonumber\\[0.3cm]
&=&-\frac{1}{2\pi i}\left\{\,\int_\infty^0(-dr)\,e^{-\xi r}(re^{i\pi})^{-\nu/2}K_{\rho}(\sqrt{8e^x}\,r^{1/2}e^{i\pi/2})\right. \nonumber\\
& &\hspace{1.5cm}\left.   +\int_0^\infty(-dr)\,e^{-\xi r}(re^{-i\pi})^{-\nu/2}K_{\rho}(\sqrt{8e^x}\,r^{1/2}e^{-i\pi/2})\,\right\}  \nonumber\\[0.2cm]
&=&-\frac{1}{2\pi i}\left\{\,e^{-i\pi\nu/2}\int_0^\infty dr\,e^{-\xi r}r^{-\nu/2}K_{\rho}(\sqrt{8e^x}\,r^{1/2}e^{i\pi/2})\right. \nonumber\\
& &\hspace{1.5cm}\left.   -\,e^{i\pi\nu/2}\int_0^\infty dr\,e^{-\xi r}r^{-\nu/2}K_{\rho}(\sqrt{8e^x}\,r^{1/2}e^{-i\pi/2})\,\right\} \nonumber 
\end{eqnarray}
\begin{eqnarray}
&=&\frac{1}{2\pi}\,\,\frac{\xi^{(\nu-1)/2}}{\sqrt{8e^x}}\,\,e^{-e^x/\xi}\,\,
   \Gamma\left(\frac{2-\nu+\rho}{2}\right)\,\Gamma\left(\frac{2-\nu-\rho}{2}\right)\nonumber\\[0.2cm]
& &\hspace{0.8cm}\times\,\left\{e^{-i\pi\nu/2}\,W_{\frac{\nu-1}{2},\frac{\rho}{2}}\left(\frac{2\,e^x}{\xi}\,e^{i\pi}\right)+
   e^{i\pi\nu/2}\,W_{\frac{\nu-1}{2},\frac{\rho}{2}}\left(\frac{2\,e^x}{\xi}\,e^{-i\pi}\right)\right\}\,. \label{L}
\end{eqnarray}
In the last line above we have used the formula (6.643.3) of \cite{GR07} to evaluate the integration in terms of
the Whittaker function $W_{\kappa,\mu}(z)$. Note that $z=0$ is the branch point of $W_{\kappa,\mu}(z)$,
so $W_{\kappa,\mu}(ze^{i\pi})\neq W_{\kappa,\mu}(ze^{-i\pi})$. To further simplify (\ref{L}), we use the relation
between $W_{\kappa,\mu}(z)$ and the other Whittaker function $M_{\kappa,\mu}(z)$
\begin{eqnarray}
W_{\kappa,\mu}(z)=\frac{\Gamma(-2\mu)}{\Gamma\left(\frac{1}{2}-\kappa-\mu\right)}\,M_{\kappa,\mu}(z)+
                  \frac{\Gamma(2\mu)}{\Gamma\left(\frac{1}{2}-\kappa+\mu\right)}\,M_{\kappa,-\mu}(z)\,,
\end{eqnarray}
and the Kummer's relation $M_{\kappa,\mu}(z\,e^{\pm i\pi})=e^{\pm i\pi(\mu+1/2)}M_{-\kappa,\mu}(z)$
to take out the minus sign, together with $\Gamma(1/2+z)\,\Gamma(1/2-z)=\pi/\cos(\pi z)$, then we have
\begin{eqnarray}
I&=&\frac{1}{\sqrt{8e^x}}\,\,\xi^{(\nu-1)/2}\,e^{-e^x/\xi}\,\,W_{\frac{1-\nu}{2},\frac{\rho}{2}}\left(\frac{2\,e^x}{\xi}\right)
\end{eqnarray}
Therefore the pricing kernel $\mathcal{P}(\tau;x,\xi)$ in (\ref{O4}) is
\begin{eqnarray}
\mathcal{P}(\tau,x,\xi)&=&\frac{1}{2\pi^2}\int_0^\infty e^{-(u^2+\nu^2)\tau/8}\,e^{-e^x/\xi}\left(\frac{2\,e^x}{\xi}\right)^{(1-\nu)/2}\nonumber\\[0.1cm]
&&\hspace{1.2cm}\times\,\, W_{\frac{1-\nu}{2},\frac{iu}{2}}\left(\frac{2\,e^x}{\xi}\right)\,
   \left|\,\Gamma\left(\frac{\nu+iu}{2}\right)\,\right|^{\,2}\sinh(\pi u)\,u\,du  \label{PK}\\[0.2cm]
&&\hspace{-1.5cm} +\,\sum_{n=0}^{[-\nu/2\,]}\frac{2(-\nu-2n)}{n!\,\Gamma(-\nu-n+1)}\,\,e^{n(\nu+n)\tau/2}\,e^{-e^x/\xi}\left(\frac{2\,e^x}{\xi}\right)^{(1-\nu)/2}
   W_{\frac{1-\nu}{2},-\frac{\nu}{2}-n}\left(\frac{2\,e^x}{\xi}\right)\,. \nonumber
\end{eqnarray}

\section{Integrating out $\xi$}

In this section we will specify payoff functions for put and call options and then study their values.

\subsection{Put options}

The payoff function for the asian put option is
\begin{eqnarray}
\phi_P(\xi)&=&(K-\xi)\,\,\theta(K-\xi)\,,
\end{eqnarray}
where $\theta(\cdot)$ is the Heaviside step function: $\theta(x)=1$ for $x\geq0$ and $\theta(x)=0$ for $x<0$.
According to (\ref{O}) its value is
\begin{eqnarray}
\mathcal{O}_P&=&\frac{e^{-R\tau}}{\,4\,\tau\, e^x}\,\int_{0}^{K\tau}\left(K\tau-\xi\right)\,\,\mathcal{P}(\tau,x,\xi)\,\,d\xi\,.
\end{eqnarray}
By investigate the expression (\ref{PK}) of the pricing kernel $\mathcal{P}(\tau,x,\xi)$,
we see that we need to consider the following type integration
\begin{eqnarray}
I_P&=&\int_{0}^{K\tau}(K\tau-\xi)\,\,\left(\frac{2\,e^x}{\xi}\right)^{(1-\nu)/2}
      \exp\left(-\frac{e^x}{\xi}\right)\,W_{\frac{1-\nu}{2},\frac{\rho}{2}}\left(\frac{2\,e^x}{\xi}\right)\,d\xi \nonumber\\[0.2cm]
&=&4\,e^{2x}(2k)^{(3+\nu)/2}\int_1^\infty(1-y)\,y^{-3+(1-\nu)/2}\,\exp\left(-\frac{y}{4k\,}\right)\,
  W_{\frac{1-\nu}{2},\frac{\rho}{2}}\left(\frac{y}{2k}\right)\,\,dy  \nonumber\\[0.2cm]
&=&4\,e^{2x}(2k)^{(3+\nu)/2}\exp\left(-\frac{1}{4k\,}\right)W_{-\frac{3+\nu}{2},\frac{\rho}{2}}\left(\frac{1}{2k}\right)\,,\label{IP}
\end{eqnarray}
where $k=K\tau/(4e^x)$, and we have use the formula (7.623.7) of \cite{GR07} in the last line.
Therefore the value of an Asian call option is
\begin{eqnarray}
\mathcal{O}_P&=&\frac{e^{R\tau+x}}{2\pi^2\tau}\int_0^\infty e^{-(u^2+\nu^2)\tau/8}
   (2k)^{(3+\nu)/2}\,e^{-1/(4k)}\,W_{-\frac{3+\nu}{2},\frac{iu}{2}}\left(\frac{1}{2k}\right) \nonumber\\
&&\hspace{1.8cm}\times\,\left|\,\Gamma\left(\frac{\nu+iu}{2}\right)\,\right|^{\,2}\sinh(\pi u)\,u\,du  \label{PO}\\[0.2cm]
&&\hspace{-1.0cm} +\,\,\frac{e^{R\tau+x}}{\tau}\,\sum_{n=0}^{[-\nu/2\,]}\frac{2(-\nu-2n)}{n!\,\Gamma(-\nu-n+1)}\,\,e^{n(\nu+n)\tau/2}
   (2k)^{(3+\nu)/2}\,e^{-1/(4k)}\,W_{-\frac{3+\nu}{2},-\frac{\nu}{2}-n}\left(\frac{1}{2k}\right)\,. \nonumber
\end{eqnarray}
By using the relation (see (9.237.3) of \cite{GR07} \footnote{The factor $n!$ is missed in \cite{GR07}.})
\begin{eqnarray}
W_{-\frac{3+\nu}{2},-\frac{\nu}{2}-n}(z)&=&(-1)^n\,n!\,z^{-n-(3+\nu)/2}\,e^{-z/2}\,L_n^{-\nu-2n}(z)
\end{eqnarray}
with $L_n^{-\nu-2k}(z)$ being the generalized Laguerre polynomial, and
\begin{eqnarray}
W_{\mu-1/2,\,\mu}(z)&=&z^{1/2-\mu}\,e^{z/2}\,\Gamma(2\mu,z)
\end{eqnarray}
with $\Gamma(2\mu,z)$ the incomplete Gamma function, it can be shown that (\ref{PO}) is exactly
equal \footnote{Due to different conventions, we need the replacement $\tau_{\mathrm{here}}=4\tau_{\mathrm{there}}$ and ${e^x}_{\mathrm{here}}=S_{0\,\mathrm{there}}$.}
to the result obtained in \cite{L04} through a different approach. \cite{L04} is based on
an equivalence between two stochastic process, while our method seems more elementary, just by doing
integrations. Actually we can derive that equivalence by using the formulation in this paper.

\subsection{Call options}

Now we consider the Asian call option, whose payoff function is
\begin{eqnarray}
\phi_C(\xi)&=&(\,\xi-K)\,\,\,\theta(\,\xi-K)\,.
\end{eqnarray}
The integration we need to do is
\begin{eqnarray}
I_C&=&\int_{K\tau}^{\infty}(\,\xi-K\tau)\,\,\left(\frac{2\,e^x}{\xi}\right)^{(1-\nu)/2}
      \exp\left(-\frac{e^x}{\xi}\right)\,W_{\frac{1-\nu}{2},\frac{\rho}{2}}\left(\frac{2\,e^x}{\xi}\right)\,d\xi \nonumber\\[0.2cm]
&=&4\,e^{2x}(2k)^{(3+\nu)/2}\int_0^1(1-y)\,y^{-3+(1-\nu)/2}\,\exp\left(-\frac{y}{4k\,}\right)\,
  W_{\frac{1-\nu}{2},\frac{\rho}{2}}\left(\frac{y}{2k}\right)\,\,dy  \nonumber\\[0.2cm]
&=&4\,e^{2x}(2k)^{(3+\nu)/2}\exp\left(-\frac{1}{4k\,}\right)W_{-\frac{3+\nu}{2},\frac{\rho}{2}}\left(\frac{1}{2k}\right)\,.\label{CO}
\end{eqnarray}
We have used the formula (6.623.8) of \cite{GR07} in the last line. Note that,
unlike (\ref{IP}) is always true, (\ref{CO}) is convergent only under the condition
$\nu+|\mathrm{Re}\,\rho|<-2$. For the integration part of the pricing kernel $\mathcal{P}(\tau,x,\xi)$
to be convergent, we should require $\nu<-2$. Then there are at least $n=0,1$ two terms in the finite
summation part of (\ref{PK}). But these two terms are both divergent since $\nu+(-\nu-2n)\geq-2$ for $n=0,1$.
Therefore for call options we cannot naively interchange the order of the integrations over $u$ and $\xi$.
However we have the so-called put-call parity relation \cite{GY93}
\begin{eqnarray}
\mathcal{O}_C=\mathcal{O}_P+\frac{1-e^{-R\tau}}{R\tau}\,\,e^x-\,e^{-R\tau}K\,.
\end{eqnarray}
We can use this relation to obtain the value of Asian call options from that of put options.

\appendix
\section{The completeness proof}

In this appendix we will prove the completeness of the family (\ref{EigFn}),  i.e. for $q>0$,
\begin{eqnarray}
I(x,x')\,:=\,\frac{1}{\pi^2}\int_0^{\infty}K_{iu}(\sqrt{8q}\,e^{x/2})\,K_{iu}(\sqrt{8q}\,e^{x'/2})\,\sinh(\pi u)\,u\,du\,=\,\delta(x-x')\,.\label{delta}
\end{eqnarray}
Since the integrand is an even function due to $K_{iu}=K_{-iu}$, we can extend the integration range to $(-\infty,\infty)$.
Therefore
\begin{eqnarray}
I(x,x')&=&\frac{i}{2\pi^2}\,\lim_{\Lambda\rightarrow\infty}\int_{-i\Lambda}^{i\Lambda}
       K_{\nu}(\sqrt{8q}\,e^{x/2})\,K_{\nu}(\sqrt{8q}\,e^{x'/2})\sin(\pi\nu)\,\nu\,d\nu\,.
\end{eqnarray}
By using the relation $K_{\nu}(z)=2^{-1}\pi(I_{-\nu}(z)-I_{\nu}(z))/\sin(\pi\nu)$,
we decompose $I(x,x')$ into three terms
\begin{eqnarray}
I(x,x')&=&\frac{i}{8}\,\lim_{\Lambda\rightarrow\infty}\left\{
         \int_{-i\Lambda}^{i\Lambda}\,\frac{\nu\,d\nu}{\sin(\pi\nu)}\,I_{\nu}(z)\,I_{\nu}(z')+
         \int_{-i\Lambda}^{i\Lambda}\,\frac{\nu\,d\nu}{\sin(\pi\nu)}\,I_{-\nu}(z)\,I_{-\nu}(z') \right.\nonumber\\[0.2cm]
&&\hspace{1.5cm}\left.-
         \int_{-i\Lambda}^{i\Lambda}\,\frac{\nu\,d\nu}{\sin(\pi\nu)}\,[\,I_{\nu}(z)\,I_{-\nu}(z')+I_{-\nu}(z)\,I_{\nu}(z')\,]\right\}\,,
\end{eqnarray}
where $z=\sqrt{8q}\,e^{x'/2}$ and $z'=\sqrt{8q}\,e^{x'/2}$. The first two terms are actually equal
by interchanging $\nu$ and $-\nu$.
Since the integrands are holomorphic in the complex $\nu$-plane, we can deform the integration path
to a semicircle $C_\Lambda:=\{\Lambda\,e^{i\phi}|-\frac{\pi}{2}\leq\phi\leq\frac{\pi}{2}\}$. Therefore
\begin{eqnarray}
I(x,x')&=&\frac{i}{8}\,\lim_{\Lambda\rightarrow\infty}\left\{\,
          2\int_{C_{\Lambda}}\frac{\nu\,d\nu}{\sin(\pi\nu)}\,I_{\nu}(z)\,I_{\nu}(z')\right. \nonumber\\[0.2cm]
&&\hspace{1.8cm}\left.
          -\int_{C_{\Lambda}}\frac{\nu\,d\nu}{\sin(\pi\nu)}\,[\,I_{\nu}(z)\,I_{-\nu}(z')+I_{-\nu}(z)\,I_{\nu}(z')\,]\,\right\}\,. \label{I}
\end{eqnarray}
When the order $\nu$ is large and $z$ is fixed, we have
\begin{eqnarray}
I_\nu(z)\,\sim\,\,\frac{1}{\,\Gamma(1+\nu)}\left(\frac{z}{2}\right)^\nu\,,\qquad
|\nu|\rightarrow\infty\,,\quad
|\mathrm{arg}z|<\pi\,.
\end{eqnarray}
By carefully analyzing the asymptotic behavior along $C_\Lambda$ when $\Lambda\rightarrow\infty$,
it can be shown that \cite{F95} the first term of (\ref{I}) actually tends to zero.
So we have
\begin{eqnarray}
I(x,x')&=&-\frac{i}{8}\,\lim_{\Lambda\rightarrow\infty}\,\int_{C_{\Lambda}}\frac{\nu\,d\nu}{\sin(\pi\nu)}\,
       \left\{\frac{\left(\frac{z}{2}\right)^{\nu}}{\,\Gamma(1+\nu)}\,\frac{\left(\frac{z'}{2}\right)^{-\nu}}{\,\Gamma(1-\nu)}+
       \frac{\left(\frac{z}{2}\right)^{-\nu}}{\,\Gamma(1-\nu)}\,\frac{\left(\frac{z'}{2}\right)^{\nu}}{\,\Gamma(1+\nu)}\right\} \nonumber\\[0.2cm]
&=&\lim_{\Lambda\rightarrow\infty}\,\,\frac{\Lambda}{8\pi}\int_{-\frac{\pi}{2}}^{\frac{\pi}{2}}\,
       \left\{\left(\frac{z}{z'}\right)^{\Lambda e^{i\phi}}+\left(\frac{z}{z'}\right)^{-\Lambda e^{i\phi}}\right\}\, e^{i\phi}\,d\phi  \nonumber\\[0.2cm]
&=&\lim_{\Lambda\rightarrow\infty}\,\,\frac{\Lambda}{8\pi}\,\sum_{n=0}^{\infty}\,
       \left\{\frac{(\frac{x-x'}{2})^n\Lambda^n}{n!}+(-1)^n\frac{(\frac{x-x'}{2})^n\Lambda^n}{n!}\right\}
       \int_{-\frac{\pi}{2}}^{\frac{\pi}{2}}e^{i(n+1)\phi}d\phi \nonumber\\[0.2cm]
&=&\frac{1}{2}\,\lim_{\Lambda\rightarrow\infty}\,\frac{\sin(\Lambda\frac{(x-x')}{2})}{\pi\,\frac{(x-x')}{2}}\,
=\,\frac{1}{2}\times\delta\left(\frac{x-x'}{2}\right)=\,\delta(x-x')\,.
\end{eqnarray}
In the second line we have used $\Gamma(1+\nu)\,\Gamma(1-\nu)=\pi\nu/\sin(\pi\nu)$, and in the last line
the limit representation of the Dirac delta function
\begin{eqnarray}
\lim_{\Lambda\rightarrow\infty}\,\frac{\sin(\Lambda u)}{\pi u}\,=\,\delta(u)\,.
\end{eqnarray}
Therefore we have proved the completeness relation (\ref{delta}).


\begin{thebibliography}{99}

\bibitem{FH65} R. P. Feynman and A. R. Hibbs, Quantum Mechanics and Path Integrals, McGraw-Hill (1965).
\bibitem{D88} J. Dash, Path integrals and options - I, CNRS Preprint CPT-88/PE.2206 (1988).
\bibitem{D89} J. Dash, Path integrals and options - II, CNRS Preprint CPT-89/PE.2333 (1989).
\bibitem{L98} V. Linetsky, The path integral approach to financial modeling and options pricing, Computational
              Economics 11: 129 (1998).
\bibitem{M00} A. Matacz, Path dependent option pricing: the path integral partial averaging method
              [arXiv:cond-mat/0005319].
\bibitem{B97} B. E. Baaquie, A path integral approach to option pricing with stochastic volatility: some exact results,
              J. Phys. I France 7: 1733 (1997) [arXiv:cond-mat/9708178].
\bibitem{B98} B. E. Baaquie, Quantum field theory of treasury bonds, Phys. Rev. E 64: 016121 (2001) [arXiv:cond-mat/9809199].

\bibitem{O98} M. Otto, Using path integrals to price interest rate derivatives, [arXiv:cond-mat/9812318].
\bibitem{BRT99} E. Bennati, M. Rosa-Clot and S. Taddei, A path integral approach to derivative security pricing: I.
              formalism and analytical results, International Journal of Theoretical and Applied Finance 2: 381 (1999) [arXiv:cond-mat/9901277].
\bibitem{RT99} M. Rosa-Clot and S. Taddei, A path integral approach to derivative security pricing: II. numerical methods,
              International Journal of Theoretical and Applied Finance 5: 123 (2002) [arXiv:cond-mat/9901279].
\bibitem{BCS02} B. E. Baaquie, C. Coriano and M. Srikant, Hamiltonian and Potentials in Derivative Pricing Models:
               Exact Results and Lattice Simulations, Physica A 334: 531 (2004) [arXiv:cond-mat/0211489].
\bibitem{BMMN04} G. Bormetti, G. Montagna, N. Moreni and O. Nicrosini, Pricing exotic options in a path integral approach,
              Quantitative Finance 6: 55 (2006) [arXiv:cond-mat/0407321].
\bibitem{LWTF08} D. Lemmens, M. Wouters, J. Tempere and S. Foulon, A path integral approach to closed-form option pricing formulas
              with applications to stochastic volatility and interest rate models, Phys. Rev. E 78: 016101 (2008) [arXiv:0806.0932].
\bibitem{DLT09} J. Devreese, D. Lemmens, J. Tempere, Path integral approach to Asian options in the Black-Scholes model,
              Physica A 389: 780 (2010).
\bibitem{D04} J. Dash, Quantitative Finance and Risk Management: A Physicist's Approach, World Scientific (2004).
\bibitem{B04} B. E. Baaquie, Quantum Finance: Path Integrals and Hamiltonians for Options and Interest Rates,
              Cambridge University Press (2004).
\bibitem{GY93} H. Geman and M. Yor, Bessel processes, Asian options and perpetuities, Mathematical Finance 3: 349 (1993).
\bibitem{L04} V. Linetsky, Spectral expansions for Asian (average price) options, Operations Research 52: 856 (2004).


\bibitem{H08} J. C. Hull, Options, Futures and Other Derivatives, 7th edition, Prentice Hall (2008).
\bibitem{DJ82} D'Hoker and R. Jackiw, Classical and quantum Liouville field theory, Phys. Rev. D26: 3517 (1982).
\bibitem{GR07} I. S. Gradshteyn and I. M. Ryzhik, Table of Integrals, Series, and Products, 7th edition,
             Academic Press (2007).
\bibitem{PP10} C. Pintoux and N. Privault, A direct solution to the Fokker-Planck equation for exponential Brownian functionals,
             Analysis and Applications 8: 287 (2010).
\bibitem{F95} T. F\"{u}l\"{o}p, Reduced $SL(2,R)$ WZNW quantum mechanics, J. Math. Phys. 37: 1617 (1996) [arXiv:hep-th/9502145].


\end{thebibliography}
\end{document}